\documentstyle[prl,aps,multicol,epsf]{revtex}

\newcommand{\CaptionI} {Schematic representation
of the models. In (a): the model with a coupling of the
superconducting site $x$
to $N$ ferromagnetic electrodes.
In (b), the model with a coupling to two ferromagnetic
electrodes.}

\newcommand{\CaptionII} {Schematic representation of the quantum switch
to probe correlated pairs of electrons.
A current source is connected to
the superconductor. In (a), there is a finite current flowing.
In (b), there is no current flowing.}

\newcommand{\CaptionIII} {Schematic representation of the
three-terminal device used to probe linear superposition of
Cooper pairs.
The insert shows the presence/absence of a
current flowing into the superconductor
as a function of the spin orientation in the ferromagnetic
reservoirs.}

\begin{document}

\draft                                          
\title{Superconducting crossed correlations in ferromagnets:\\
implications for thermodynamics and quantum transport
}

\author{R. M\'elin}
\address{Centre de Recherches sur les Tr\`es Basses
Temp\'eratures (CRTBT)\\
CNRS, BP 166 X, 38042 Grenoble Cedex, France
{}\\
and
{}\\
Laboratoire de Physique, Ecole Normale Sup\'erieure de Lyon\\
46 All\'ee d'Italie, 69364 Lyon Cedex 07, France}

\date{\today}

\maketitle
\begin{abstract}
It is demonstrated that non local Cooper pairs
can propagate in ferromagnetic electrodes
having an opposite spin orientation. 
In the presence of such crossed correlations, the
superconducting gap is found to depend explicitly
on the relative orientation of the ferromagnetic electrodes.
Non local Cooper pairs can in principle be 
probed with dc-transport.
With two ferromagnetic electrodes,
we propose a ``quantum switch'' that can be used to
detect correlated pairs of electrons.
With three or more ferromagnetic electrodes, the
Cooper pair-like state is a linear superposition
of Cooper pairs which could be detected
in dc-transport.
The effect also induces an enhancement of
the ferromagnetic proximity effect on the
basis of crossed superconducting correlations
propagating along domain walls.
\end{abstract}
\pacs{PACS numbers: 74.50.+r, 03.67.-a, 74.80.Fp}

\begin{multicols}{2}

Ferromagnetism and superconductivity are antagonist correlated
states of matter. In ferromagnetism, one spin population
is favored because of spin symmetry breaking, while in
$s$-wave superconductivity, electrons with an opposite spin
are bound into Cooper pairs because of the
attractive electron-electron interaction.
It has been a long standing problem to determine
to what extend these two orders can coexist in the same
system. As first proposed 40 years ago by Anderson and Suhl,
the coexistence is possible if the ferromagnet
acquires a cryptomagnetic~\cite{Anderson},
or cryptomagnetic-like~\cite{Buzdin} domain structure.
On the other hand, in superconductor~/ ferromagnet
heterostructures, a Cooper pair penetrating into a 
{\sl single
domain} ferromagnet acquires a finite kinetic energy
due to the coupling to the exchange field.
This results in a spatial
oscillation of the induced superconducting order
parameter~\cite{Fulde,Larkin,Clogston,Demler},
giving rise to the so-called
$\pi$-state, which has been probed
recently in two experiments~\cite{Ryazanov,Kontos}.
In this Letter, we consider Cooper pair penetration
in a {\sl multi domain} ferromagnet. It has been already
shown theoretically that crossed Andreev reflections can arise 
in a heterostructure in which two ferromagnets with
an opposite spin orientation are connected to
a superconductor~\cite{Feinberg}. Such Andreev reflections
do not exist when a single domain ferromagnet is
in contact with a superconductor~\cite{deJong,Soulen,Upadhyay}.
We demonstrate here that 
quasi long range superconducting correlations
can propagate in two magnetic domains with an opposite
magnetization.
These
correlations correspond to non local Cooper pair-like
objects in which the spin-up (down) electron propagates
in a spin-up (down) ferromagnetic domain.

This implies several consequences that
may be tested in future experiments. First,
considering the problem
from the point of view of a superconductor
order parameter
coupled to a ferromagnetic environment,
we show that
the transition temperature of the superconductor
depends explicitly on the relative spin orientations
of the electrodes. The superconducting gap is
smaller when
the electrodes are misoriented. 

The second implication of the model
is that ferromagnetic domain walls can propagate
superconducting crossed correlations, in which the
two electrons making a Cooper pair reside in
neighboring magnetic domains.
This may apply to explain the enhancement of
the proximity effect observed in
ferromagnet~/ superconductor
heterostructures~\cite{Giroud,Petrashov,Chandra,Bauer}.

The third implication of the model is related to
the production and measurement of linear superpositions
of non local Cooper pairs.
It was stressed by  Einstein, Podolsky and Rosen (EPR)
in 1935~\cite{EPR} that non locality was a deep
feature of quantum mechanics. Non locality~\cite{Bell}
has been probed experimentally
with photons~\cite{Aspect,Kwiat}.
Condensed matter systems provide maybe the opportunity to
fabricate entangled states with electrons, being massive 
particles, and to fabricate
quantum bits, which would be the building blocks
of a quantum computer~\cite{Schor,Grover,Steane,Nakamura}. 
Two proposals have been made
recently: one
is based on tunneling in a double quantum dot~\cite{Loss},
and the other is based on noise correlations of Cooper
pairs emitted in a beam splitter~\cite{Martin}.
We show that
superconducting crossed correlations in ferromagnets
provide a 
possibility to manipulate linear superpositions
of Cooper pairs.
With two ferromagnetic electrodes,
we propose a ``quantum switch'' device that
can be used to detect correlated pairs of electrons.
Linear superpositions can be obtained
with three or more ferromagnetic electrodes,
and can be probed in a dc-transport.

\begin{figure}[tbp] 
\centerline{\epsfxsize=7.5cm \epsfbox{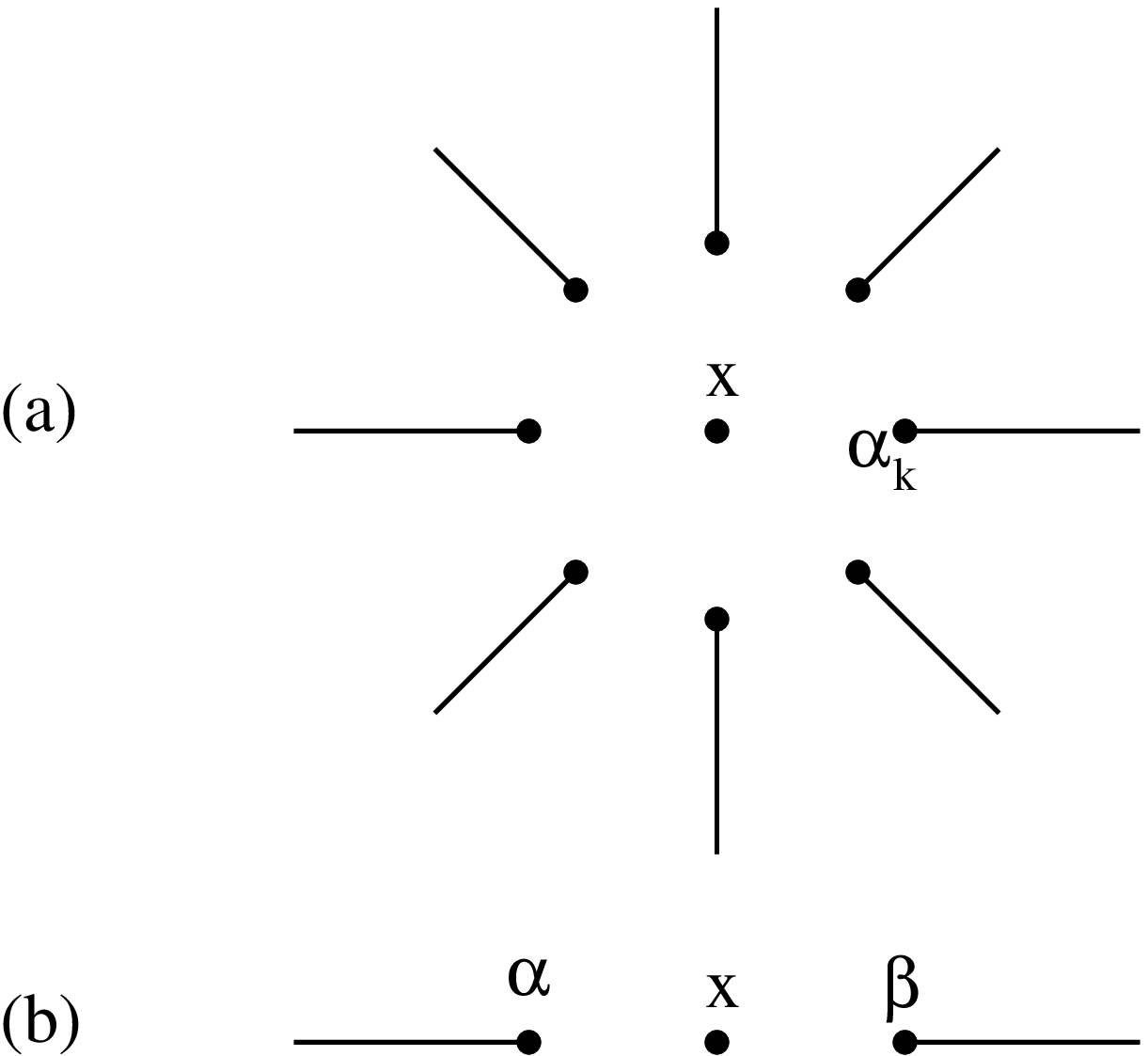}}
\bigskip
\caption{\CaptionI}
\label{fig:fig1} 
\end{figure}

Let us now consider a microscopic model in which
a superconductor is connected to external electrodes.
The superconductor is represented by the single site
effective Nambu
Green's function~\cite{Cuevas}
$\hat{g}^{R,A}(\omega) = g(\omega \pm i \eta) \hat{I}
+ f(\omega \pm i \eta) \hat{\sigma}^x$,
with
$g(\omega) =- \pi \rho_N \omega /
\sqrt{ \Delta^2 - \omega^2}$, and 
$f(\omega) = \pi \rho_N \Delta/
\sqrt{ \Delta^2 - \omega^2}
$, and
where $\rho_N$, having the dimension of an
inverse energy,
is the normal state
density of states, $\hat{I}$ is the $2 \times 2$
identity matrix, and $\hat{\sigma}^x$ a
Pauli matrix.
We assume that $N$ ferromagnetic electrodes are
in contact with the superconductor (see Fig.~\ref{fig:fig1}~(a)),
with a hopping
Hamiltonian $W = \sum_{k=1}^N t_{x,\alpha_k}
\left[ c_{\alpha_k}^+ c_x + 
c_x^+ c_{\alpha_k} \right]$.
The electrode $k$
having a spin polarization
$P_k = (\rho_{k, \uparrow} - \rho_{k,\downarrow}) /
(\rho_{k, \uparrow} + \rho_{k,\downarrow})$
is represented by the Green's function
$
\hat{g}^{A,R}_k = \pm i \pi
\left[ \rho_{k,\uparrow} ( \hat{I} +
\hat{\sigma}^z)/2 +
\rho_{k,\downarrow} ( \hat{I} -
\hat{\sigma}^z)/2 \right]
$. 
We use a perturbation theory in the
tunnel amplitude $W$, which we sum up to infinite
order~\cite{Cuevas,Nozieres}. The Dyson equation takes the form
\begin{equation}
\label{eq:Dyson}
\hat{G}_{x,x}^{R,A}= \left[ \hat{I} - 
\sum_{k=1}^N
\hat{g}_{x,x}^{R,A}
\hat{t}_{x,\alpha_k}
\hat{g}_{\alpha_k,\alpha_k}^{R,A}
\hat{t}_{\alpha_k,x} \right]^{-1}
\hat{g}_{x,x}^{R,A}
,
\end{equation}
where $\hat{t}_{\alpha_k,x}$ is the Nambu representation
of the tunnel matrix element:
$\hat{t}_{\alpha_k,x} = t_{\alpha_k,x}
\hat{\sigma}^z$. The relevant parameters appear to be
the spectral line-width associated to spin-$\sigma$
electrons: $\Gamma_\sigma = \sum_{k=1}^N
\Gamma_{k,\sigma}$, with
$\Gamma_{k,\sigma} = (t_{\alpha_k,x})^2
\rho_{k,\sigma}$. Solving Eq.~\ref{eq:Dyson} leads to
\begin{eqnarray}
\label{eq:solu-G}
\hat{G}_{x,x}^A &=& \frac{1}{\cal D}
\left\{ g \hat{I} + f \hat{\sigma}^x \right.\\
&+& \left.
i \pi (f^2 - g^2) 
\left[
\frac{\Gamma_\downarrow}{2}
\left( \hat{I} + \hat{\sigma}^z \right)
+ \frac{\Gamma_\uparrow}{2}
\left( \hat{I} - \hat{\sigma}^z \right) \right]
 \right\}
\nonumber
,
\end{eqnarray}
with ${\cal D} = 1 - i \pi g(\Gamma_\uparrow
+ \Gamma_\downarrow) + \pi^2 (f^2 - g^2)
\Gamma_\uparrow \Gamma_\downarrow$.
To calculate the superconducting order parameter, we need
to
solve the Dyson-Keldysh equation
$\hat{G}^{+ -} =
(\hat{I} + \hat{G}^R \otimes \hat{W} )
\otimes \hat{g}^{+ -} \otimes ( \hat{I}
+ \hat{W} \otimes \hat{G}^A )$, where
the convolution includes a sum over the labels
$x$ and $\alpha_k$.
Noting $X_\sigma = (1 - i \pi g \Gamma_{- \sigma})/{\cal D}$,
$Y_\sigma = i \pi f \Gamma_\sigma / {\cal D}$,
and using Eq.~\ref{eq:solu-G},
we obtain the exact expression of the Nambu component
of the Keldysh Green's function:
\begin{eqnarray}
\label{eq:solu-Gpm}
{}\left[ G_{x,x}^{+ -} \right]_{2,1} && =
2 i \pi n_F(\omega) \times \\
\nonumber
&{}&\left\{ \rho_g \left( X_\uparrow \overline{Y}_{\uparrow}
+ Y_{\downarrow} \overline{X}_{\downarrow} \right)
+ \rho_f \left( X_\uparrow
\overline{X}_\downarrow + Y_\downarrow
\overline{Y}_\uparrow \right) \right.\\
\nonumber
&{}&\left. + \frac{1}{\pi^2 \Gamma^\downarrow} 
\overline{Y}^\downarrow \left( X^\uparrow -1 \right)
+ \frac{1}{\pi^2 \Gamma^\uparrow} 
Y^\uparrow \left( \overline{X}^\downarrow -1 \right)
\right\}
,
\end{eqnarray}
where $n_F(\omega)$ is the Fermi distribution,
and we used the notation $\hat{\rho}
= \rho_g \hat{I} + \rho_f \hat{\sigma}^x
= \mbox{Im}[\hat{g^A}]/\pi$.
The superconducting gap is
obtained by imposing the self-consistent equation
$\Delta = U \int_{- \infty}^{+ \infty}
d \omega / (2 i \pi) [ \hat{G}^{+ -}(\omega)]_{2,1}$~\cite{Martin-Rodeiro},
with $U$ the microscopic
attractive interaction. The dominant contribution arises from
the large-$|\omega|$ behavior and we obtain
a BCS-type relation:
\begin{equation}
\label{eq:gap1}
\Delta = D \exp{ \left[ -  \frac{1}{\rho_N U}
\left( 1 + \pi \rho_N \Gamma_\uparrow \right)
\left( 1 + \pi \rho_N \Gamma_\downarrow \right)
\right]}
,
\end{equation}
with $D$ the bandwidth of the superconductor.
As an example, we consider a coupling to two ferromagnets.
With a parallel alignment of the magnetization in the
electrodes, we have $\Gamma_\uparrow = 2 \gamma$
and $\Gamma_\downarrow = 0$. With an antiparallel
alignment, we have $\Gamma_\uparrow = \Gamma_\downarrow
= \gamma$. The ratio of the two gaps is found to be
\begin{equation}
\label{eq:gap2}
\frac{\Delta_{AP}}{\Delta_P} = \exp{ \left( -
\frac{\pi^2 \rho_N \gamma^2}
{ U } \right)}
,
\end{equation}
which shows that the spin polarized
environment generates
a reduction of the superconducting gap that
depends explicitly on the spin orientation
of the environment.
The transition temperature
of the superconductor is larger if the electrodes
are in an antiparallel alignment.
This behavior should be contrasted
with another model proposed
recently~\cite{Baladie}.

As we show now,
the gap variation Eq.~\ref{eq:gap2} is due to the possibility
that superconducting pairs can delocalize in the
ferromagnetic electrodes having an opposite spin orientation.
Let us consider the problem with two electrodes only.
The two electrodes are labeled by the Greek indices
$\alpha_1=\alpha$ and $\alpha_2=\beta$.
We use Eq.~\ref{eq:solu-Gpm} to calculate exactly
the crossed Keldysh Green's functions:
\begin{eqnarray}
\label{eq:Gpm1}
\left[ G_{\alpha,\beta}^{+ -} \right]_{2,1} &=&
i \langle c_{\beta,\uparrow}^+ c_{\alpha,\downarrow}^+ \rangle
= \pi^2 t_\alpha t_\beta \rho_{\alpha,\downarrow}
\rho_{\beta,\uparrow} \left[ G_{x,x}^{+ -} \right]_{2,1}\\
\label{eq:Gpm2}
\left[ G_{\alpha,\beta}^{+ -} \right]_{1,2} &=&
i \langle c_{\beta,\downarrow} c_{\alpha,\uparrow} \rangle
= \pi^2 t_\alpha t_\beta \rho_{\alpha,\uparrow}
\rho_{\beta,\downarrow} \left[ G_{x,x}^{+ -} \right]_{1,2}
,
\end{eqnarray}
with $\left[ \hat{G}_{x,x}^{+ -} \right]_{2,1} = 
\overline{\left[ \hat{G}_{x,x}^{+ -} \right]}_{1,2}$
given by Eq.~\ref{eq:solu-Gpm}. 
The density of states prefactors
in Eqs.~\ref{eq:Gpm1},~\ref{eq:Gpm2} appear to be a direct
consequence of the Pauli exclusion principle. To
show this, we consider Eqs.~\ref{eq:Gpm1},~\ref{eq:Gpm2}
in the limit of fully polarized ferromagnets.
In the parallel alignment ($\rho_{\alpha,\uparrow}
= \rho_{\beta,\uparrow} =1$, $\rho_{\alpha,\downarrow}
= \rho_{\beta,\downarrow} = 0$), all pair
correlations are vanishing:
$\langle c_{\beta, \uparrow}^+ c_{\alpha,\downarrow}^+
\rangle = 
\langle c_{\beta, \downarrow} c_{\alpha,\uparrow}
\rangle = 0$. This is fully expected because one cannot
add or destroy a spin-down electron
in the presence of a spin-up band only.
For the same reason, one has
$\langle c_{\beta,\uparrow}^+ c_{\alpha,\downarrow}^+
\rangle = 0$ in the antiparallel alignment
($\rho_{\alpha,\uparrow}
= \rho_{\beta,\downarrow} =1$, $\rho_{\alpha,\downarrow}
= \rho_{\beta,\uparrow} = 0$).
The remaining non vanishing crossed correlations
are $\langle c_{\beta,\downarrow} c_{\alpha,\uparrow}
\rangle$ and $\langle c_{\beta,\downarrow}^+ 
c_{\alpha,\uparrow}^+ \rangle$. This shows the possibility
to generate superconducting crossed correlations
in two ferromagnets with an opposite
magnetization. 
To characterize the propagation of crossed correlations,
we calculate
the Gorkov function $\hat{G}^{+ -}_{i,j}$, with
$i$ and $j$ two sites in the ferromagnetic electrodes
$\alpha$  and $\beta$ such that $x_i = - x_j$. 
Assuming that
the ferromagnetic electrodes behave like a three
dimensional metal, we find
$$
\label{eq:longd1}
\left[ \hat{G}_{i,j}^{+ -} \right]_{1,2} \sim
\frac{1}{|x_i|} \pi^2 t_\alpha t_\beta \rho_{\alpha,\uparrow}
\rho_{\beta,\downarrow} \left[
\hat{G}_{x,x}^{+ -} \right]_{1,2}
.
$$
By comparison, there is a density
of states prefactor $\rho_{\alpha,\uparrow}
\rho_{\alpha,\downarrow}$ in the
{\sl local} superconducting
correlation in electrode $\alpha$.
As a consequence,
in strongly spin polarized ferromagnets,
superconducting
crossed correlations can propagate while ordinary
superconducting correlations cannot propagate.
It is well known that there is an
oscillating induced order parameter associated
to Cooper pair penetration in partially
spin polarized
ferromagnets~\cite{Buzdin,Fulde,Larkin,Clogston,Demler}.
There are no such oscillations in the case of
crossed correlations because Cooper
pairs do not acquire a center of mass
momentum when entering the ferromagnetic electrodes.

\begin{figure}[tbp] 
\centerline{\epsfxsize=7.5cm \epsfbox{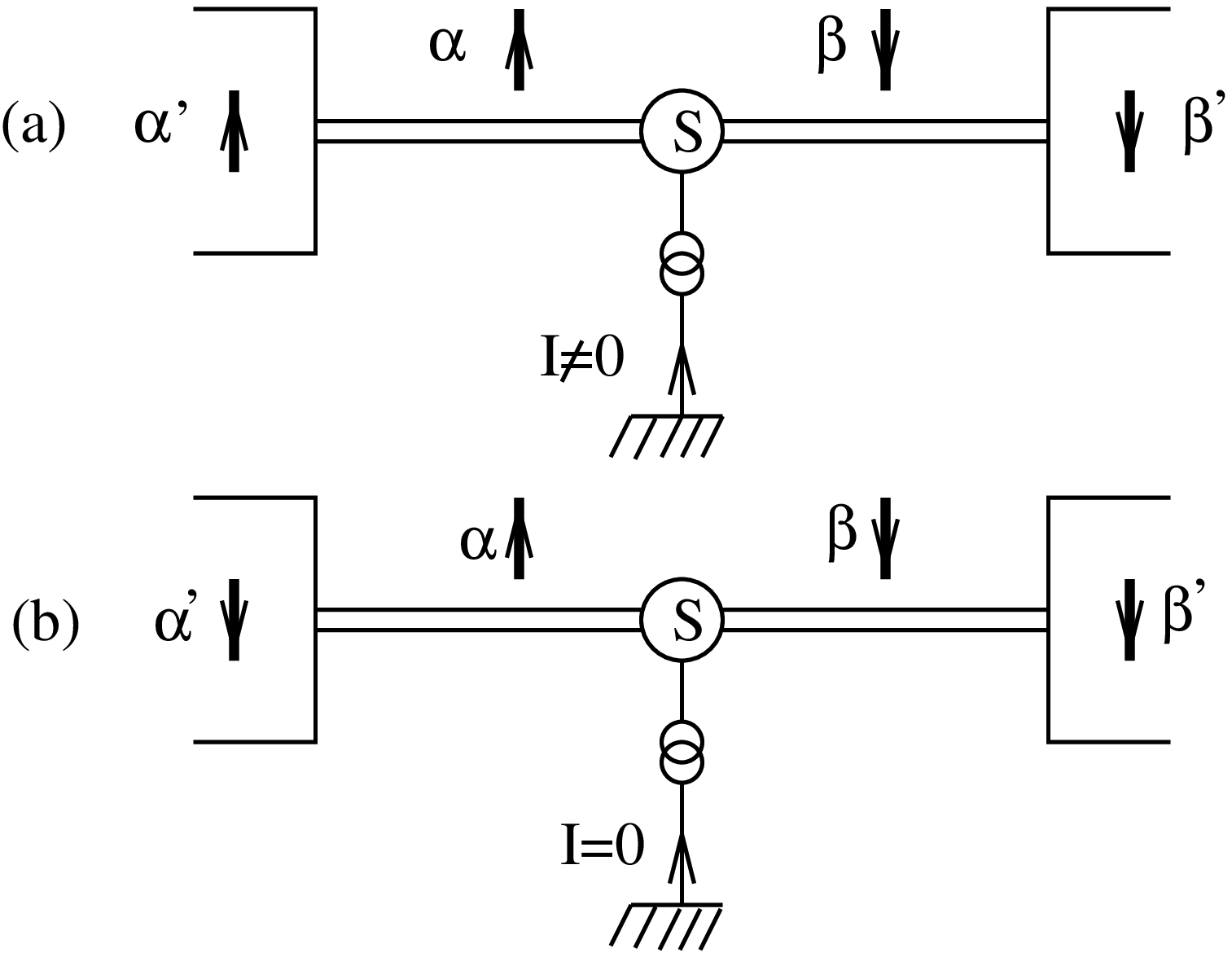}}
\bigskip
\caption{\CaptionII}
\label{fig:fig2} 
\end{figure}

The model can be considered from the point of view
of propagation of cross-correlated Cooper pairs
along domain walls in a multi domain ferromagnet.
Such crossed correlations
can generate an enhancement of the
ferromagnetic superconducting
proximity effect, which is not against recent experiments
in ferromagnet~/ superconductor
heterostructures~\cite{Giroud,Petrashov,Chandra}. 
Another proposal based on spin accumulation has been
made recently~\cite{Bauer}, but appears to be incompatible with
some experiments~\cite{Petrashov}. Our scenario and
the spin accumulation picture both contribute
to the same effect, but in a different situation:
crossed correlations can propagate only in
multi domain ferromagnets, while the spin accumulation
mechanism is valid even with single domain ferromagnets.

\begin{figure}[tbp] 
\centerline{\epsfxsize=7.5cm \epsfbox{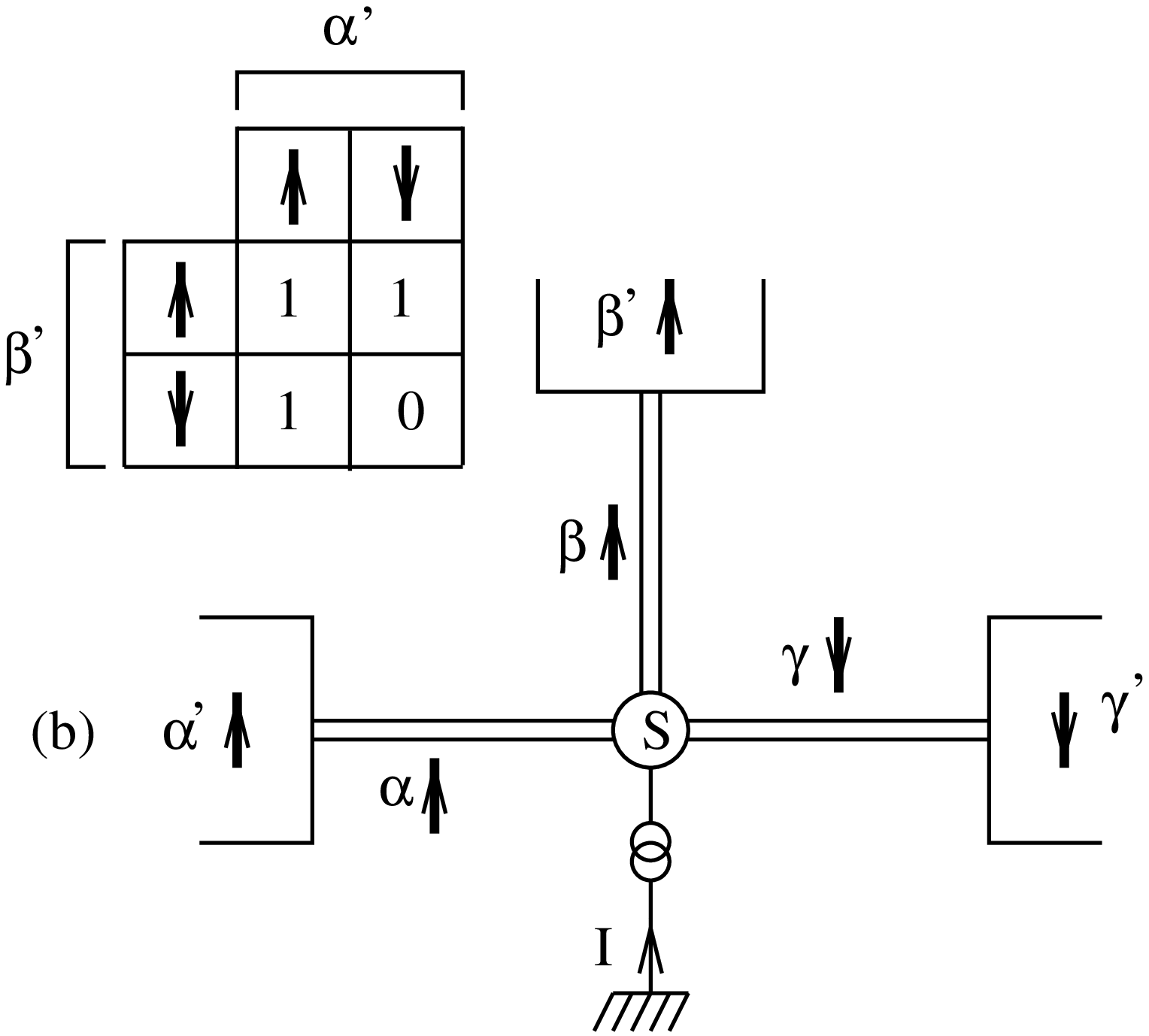}}
\bigskip
\caption{\CaptionIII}
\label{fig:fig3} 
\end{figure}

Now we show that superconducting crossed correlations
can be used to produce correlated
pairs of electrons.
Let us consider two
ferromagnets $\alpha$ and $\beta$
in contact with a superconductor.
The cross-correlated degrees of freedom are represented
by the Cooper pair-like wave function
$
| \psi \rangle = \left[ u_0 + v_0 c_{\alpha,\uparrow}^+
c_{\beta,\downarrow}^+ \right]| 0 \rangle$,
with $u_0$ and $v_0$ the BCS coherence factors.
Let us consider two additional
ferromagnetic electrodes $\alpha'$ and $\beta'$
having a spin orientation $\Sigma_{\alpha'}$ and
$\Sigma_{\beta'}$ connected to the electrodes
$\alpha$ and $\beta$ (see Fig.~\ref{fig:fig2}).
The electrodes $\alpha'$ and $\beta'$ are
considered to be reservoirs
in which all inelastic processes take place.
For the sake of obtaining the basic physics of
such systems, we restrict ourselves to
fully polarized ferromagnets and high transparency
contacts~\cite{note-bis}.
If
$\Sigma_{\alpha'} = \uparrow$, $\Sigma_{\beta'} = \downarrow$,
the correlated pair
can be transmitted into the reservoirs $\alpha'$
and $\beta'$ and a finite current is flowing into the
superconductor (see Fig.~\ref{fig:fig2}~(a)).
If $\Sigma_{\alpha'}=\Sigma_{\beta'}=
\downarrow$, the spin-up electron making the
correlated state
is backscattered at the interface with the spin-down
ferromagnet $\alpha'$. Coming back onto the superconductor
interface it undergoes a crossed Andreev reflection~\cite{Feinberg}
in which a Cooper is formed in the superconductor and
a spin-down hole is transferred into electrode $\beta$.
The whole process does not carry electrical charge:
there is no current
transmitted into the superconductor (see Fig.~\ref{fig:fig2}~(b)).
The ``quantum switch'' device on Fig.~\ref{fig:fig2}
can therefore be used to produce
and detect correlated pairs of electrons
with dc-transport.

Now we discuss the production of linear superpositions
in a three-terminal device (see Fig.~\ref{fig:fig3}).
The three ferromagnetic electrodes are labeled
by the indices $\alpha_1=\alpha$,
$\alpha_2=\beta$ and $\alpha_3=\gamma$.
With fully polarized ferromagnets having a spin
orientation $\sigma_\alpha=\sigma_\beta=\uparrow$,
$\sigma_\gamma=\downarrow$,
the exact form of the crossed-correlations is
given by
$\left[ \hat{G}_{\alpha (\beta),\gamma}^{+ -} \right]_{1,2}
= \pi^2 t_{\alpha (\beta)} t_\gamma
\left[ \hat{G}_{x,x}^{+ -} \right]_{1,2}$. 
This means that Cooper pairs can
delocalize over several electrodes. The corresponding wave
function is a linear superposition of Cooper pairs
$| \psi \rangle =  
\lambda_\alpha \left[ u_0 + v_0 c_{\alpha, \uparrow}^+
c_{\gamma,\downarrow}^+\right]| 0 \rangle
+\lambda_\beta \left[ u_0 + v_0 c_{\beta, \uparrow}^+
c_{\gamma,\downarrow}^+ \right]| 0 \rangle$.
The coefficients $\lambda_\alpha$ and $\lambda_\beta$
are such that the Cooper pair wave function contains
the same pair correlations as the Gorkov function:
${ \langle c_{\beta,\uparrow}^+ c_{\gamma_\downarrow}^+ \rangle}
/{\langle c_{\alpha,\uparrow}^+ c_{\gamma_\downarrow}^+ \rangle}
= {\lambda_\beta}/{\lambda_\alpha} = {t_\beta}/{t_\alpha}
$,
from what we deduce 
$
\lambda_{\alpha (\beta)} =
{ t_{\alpha (\beta)}}/
{\sqrt{t_\alpha^2 + t_\beta^2 + 2 u_0^2 t_\alpha 
t_\beta}}
$.
As a direct consequence of the linear superposition,
the current
flowing into the superconductor is vanishing if
$\Sigma_{\alpha '} = \Sigma_{\beta '}
= \downarrow$ and finite in the three other
spin orientations (see Fig.~\ref{fig:fig3}).
Now the linear superposition associated to the
electrodes magnetization $\sigma_\alpha=\uparrow,
\sigma_\beta=\sigma_\gamma=\downarrow$ is
$| \psi \rangle =  
\lambda'_\beta \left[ u_0 + v_0 c_{\alpha, \uparrow}^+
c_{\beta,\downarrow}^+\right]| 0 \rangle
+\lambda'_\gamma \left[ u_0 + v_0 c_{\alpha, \uparrow}^+
c_{\gamma,\downarrow}^+ \right]| 0 \rangle$, with
$
\lambda'_{\beta (\gamma)} =
{ t_{\beta (\gamma)}}/
{\sqrt{t_\beta^2 + t_\gamma^2 + 2 u_0^2 t_\beta 
t_\gamma}}
$.
The current flowing into the superconductor is
vanishing in the two spin orientations
$\Sigma_{\alpha'}=\downarrow$,
$\Sigma_{\beta'}=\uparrow,\downarrow$
and finite otherwise. Therefore,
a dc-current measurement can make the distinction
between the linear superpositions
associated to the spin orientations
$\sigma_\alpha=\sigma_\beta=\uparrow$,
$\sigma_\gamma=\downarrow$ and
$\sigma_\alpha=\uparrow$,
$\sigma_\beta=\sigma_\gamma=\downarrow$.

To conclude, we have shown that quasi long range superconducting
crossed correlations can propagate in ferromagnets
having an opposite spin orientation. The superconducting
crossed correlations are much stronger than the local ones.
Such crossed correlations can propagate along ferromagnetic
domain walls, and contribute to an enhancement of the
ferromagnetic proximity effect, which may apply to recent
experiments~\cite{Giroud,Petrashov,Chandra}.
The superconducting gap 
depends explicitly
on the spin orientation of the ferromagnetic electrodes,
which could be used as an experimental probe
of superconducting crossed correlations.
We have shown that crossed correlations can be used to
produce correlated pairs of electrons and linear superpositions
of correlated pairs.
Such states can in principle
be detected with dc-transport.
The microscopic calculation of the current
will be the subject of a future work.

The author acknowledges unvaluable discussions with
P. Degiovanni, D. Feinberg and M. Giroud. 





\end{multicols}

\begin{references}

\bibitem{Anderson} P. Anderson and H. Suhl,
Phys. Rev. {\bf 116}, 6739 (1959).

\bibitem{Buzdin} A. Buzdin and L. Bulaevskii,
JETP {\bf 67}, 576 (1988).

\bibitem{Fulde} P. Fulde and A. Ferrel, Phys. Rev.
{\bf 135}, A550 (1964).

\bibitem{Larkin} A. Larkin and Y. Ovchinnikov,
Sov. Phys. JETP {\bf 20}, 762 (1965).

\bibitem{Clogston} M. A. Clogston, Phys. Rev. Lett. {\bf 9}, 266
(1962).

\bibitem{Demler} E.A. Demler, G.B. Arnold and M.R. Beasley,
Phys. Rev. B {\bf 55}, 15174 (1997).

\bibitem{Ryazanov} V.V. Ryazanov {\sl et al.},
arXiv:cond-mat/0008364

\bibitem{Kontos} T. Kontos, M. Aprili, J. Lesueur
and X. Grison, arXiv:cond-mat/0009192.

\bibitem{Feinberg} G. Deutscher and D. Feinberg,
App. Phys. Lett. {\bf 76},
487 (2000).

\bibitem{deJong} M.J.M.~de~Jong and~C.W.J.~Beenakker,
Phys. Rev. Lett. {\bf 74}, 1657 (1995).

\bibitem{Soulen} R.J.~Soulen {\sl et al.},
Science {\bf 282}, 85 (1998).

\bibitem{Upadhyay} S.K.~Upadhyay {\sl et al.},
Phys. Rev. Lett. {\bf 81}, 3247 (1998).

\bibitem{Giroud} M. Giroud {\sl et al.},
Phys. Rev. B {\bf 58}, R11872 (1998).

\bibitem{Petrashov} V.T. Petrashov, I.A. Sosnin,
and C. Troadec, arXiv:cond-mat/0007278.

\bibitem{Chandra} J. Aumentado and V. Chandrasekhar,
arXiv:cond-mat/0007433.

\bibitem{Bauer} W. Belzig, A. Brataas, Yu. V. Nazarov,
and G.E. Bauer, arXiv:cond-mat/0005188.

\bibitem{EPR} A. Einstein, B. Podolsky, and N. Rosen,
Phys. Rev. {\bf 47}, 777 (1935).

\bibitem{Bell} J.S. Bell, Physics (N.Y.) {\bf 1}, 195 (1965).

\bibitem{Aspect} A. Aspect, P. Grangier, and
G. Roger, Phys. Rev. Lett. {\bf 47}, 460 (1981);
A. Aspect, J. Dalibard, and G. Roger,
{\sl ibid.} {\bf 49}, 1904 (1982).

\bibitem{Kwiat} P.G. Kwiat {\sl et al.},
Phys. Rev. Lett. {\bf 75}, 4337 (1995).

\bibitem{Schor} P.W. Schor, in {\sl Proc. 35th Symposium
on the Foundations of Computer Science}
(IEEE Computer Society Press), 124 (1994).

\bibitem{Grover} L.K. Grover, Phys. Rev. Lett. {\bf 79}, 325
(1997).

\bibitem{Steane} A. Steane, Rep. Prog. Phys. {\bf 61}, 117
(1998).

\bibitem{Nakamura} Y. Nakamura, Yu. A. Pashkin, and
J.S. Tsai, Nature {\bf 398}, 786 (1999).

\bibitem{Loss} D. Loss and E.V. Sukhorukov,
Phys. Rev. Lett. {\bf 84},
1035 (2000).

\bibitem{Martin}G.B. Lesovik, T. Martin, and G.
Blatter, arXiv:cond-mat/0009193.

\bibitem{Cuevas} J.C. Cuevas, A. Martin-Rodero, and A.
Levy Yeyati, Phys. Rev. B {\bf 54}, 7366 (1996).

\bibitem{Nozieres} C. Caroli, R. Combescot, P. Nozi\`eres,
and D. Saint-James, J. Phys. C {\bf 5}, 21 (1972).

\bibitem{Martin-Rodeiro} A. Martin-Rodeiro,
F.J. Garcia-Vidal, and A.
Levy-Yeyati, Phys. Rev. Lett.
{\bf 72}, 554 (1994).

\bibitem{Baladie} I. Baladie, A. Buzdin, N. Ryzhanova,
A. Vedyayev, Phys. Rev. B {\bf 63}, 054518 (2001).

\bibitem{note-bis} The Keldysh formalism can be used to
calculate the current-voltage characteristics in the presence
of arbitrary spin polarizations and contact transparencies
[R. M\'elin, unpublished].


\end{references}
\end{document}